\documentclass[12pt,preprint]{aastex}
\usepackage{emulateapj5}
\usepackage{onecolfloat}

\newcommand{\sm}{\, {\rm M}_{\odot}}

\newcommand{\kms}{\, {\rm km~s}^{-1}}
\def\gsim { \lower .75ex \hbox{$\sim$} \llap{\raise .27ex \hbox{$>$}} }
\def\lsim { \lower .75ex \hbox{$\sim$} \llap{\raise .27ex \hbox{$<$}} }

\shorttitle{Extragalactic origin of the Arcturus group}
\shortauthors{Navarro, Helmi \& Freeman}

\begin{document}

\twocolumn[

\title{The Extragalactic Origin of the Arcturus Group}

\author{Julio F. Navarro\altaffilmark{1}}
\affil{Department of Physics and Astronomy, University of Victoria, 
Victoria, BC V8P 1A1, Canada}

\author{Amina Helmi\altaffilmark{2},}
\affil{Kapteyn Institute, P.O.Box 800, 9700 AV Groningen, The Netherlands}

\author{Kenneth C. Freeman} 
\affil{ Research School of Astronomy \&
Astrophysics, Mount Stromlo Observatory, The Australian National
University, Weston Creek ACT 2611, Australia}

\begin{abstract}
We reanalyze the group of stars associated kinematically with Arcturus
(Eggen 1971), and confirm that they constitute a peculiar grouping of
metal-poor stars with similar apocentric radius, common angular
momentum, and distinct metal abundance patterns. These properties are
consistent with those expected for a group of stars originating from
the debris of a disrupted satellite. The Arcturus group stands out
clearly in the catalog of Beers et al. (2000) as well as in other
compilations of metal-poor stars in the solar neighborhood.  Its
angular momentum appears too low to arise from dynamical perturbations
induced by the Galactic bar, and coincides with that of the
kinematically-peculiar population of stars identified above and below
the Galactic plane by Gilmore, Wyse \& Norris (2002): the Arcturus
group is in all likelihood the solar neighborhood extension of such
population. Further analysis is needed to confirm our proposal that
Arcturus, together with some of the brightest stars in the night sky,
may have formed beyond the confines of the Galaxy. However, validating
the extragalactic origin of the Arcturus group would lend support to
the prevalence of accretion events envisioned in hierarchical models
of galaxy formation, and would suggest that early mergers may have
contributed a substantial fraction of the old and metal-deficient
stars in the Galactic disk.
\end{abstract}

]
\altaffiltext{1}{Fellow of CIAR and of the J.S.Guggenheim Memorial Foundation}
\altaffiltext{2}{NOVA Fellow}

\section{Introduction}
\label{sec:intro}

The discovery of substructure in the distribution of stars in the outer regions
of galaxies like the Milky Way and M31 constitutes indisputable evidence that
accretion events and mergers have played a role in building up their stellar
halos (Ibata et al. 1994, Helmi et al. 1999, Majewski et al. 2003). Although the
role of such events in the formation of the Galactic disk components is not yet
fully understood, models of the thick disk component typically envision the
tidal heating of an extant thin disk by a merging satellite (Quinn \& Goodman
1986). Recent numerical simulations of disk galaxy formation also suggest that
disrupted satellites might have contributed a significant fraction of the old
stars in the disk of the Galaxy (Abadi et al. 2003a,b). As with the stellar
halo, long-lived traces of such events may be preserved in the form of
substructure in the orbital distribution of stars in the disk (Helmi et al
2003).

Identifying such debris ought to be easier in samples where contamination by the
thin disk component is minimized, such as samples of metal poor/old stars, or
else samples collected {\it in situ} away from the disk or far from the Galactic
center. The discovery of a ``ring'' of stars of almost certain extragalactic
origin in the outer disk of the Milky Way by the SDSS (Newberg et al 2002, Yanny
et al 2003), as well as recent report of a galaxy being cannibalized in the
Milky Way disk (Martin et al 2003) provide examples that this process indeed
seems to be at work in the Galaxy (Helmi et al 2003, Rocha-Pinto et al 2003).

Gilmore, Wyse \& Norris (2002) have also recently argued for
substructure in the Galactic thick disk. They postulate the existence
of a distinct population of stars, with kinematics intermediate
between the canonical thick disk and the stellar halo (mean rotational
velocity of $\sim 100 \kms$) to explain the results of their
spectroscopic survey of F--G stars located 0.5--5 kpc away from the
Galactic plane.

Substructure in the disk has been noted since the early days of
Galactic structure research (see, e.g., Eggen 1998 and references
therein), and has usually been interpreted as signature of the gradual
dissolution of loosely bound star clusters in the clumpy potential of
the disk.  One example is the group of stars associated by Eggen
(1971) with the bright star Arcturus, whose rotation speed lags the
Local Standard of Rest (LSR) by roughly $\sim 100 \kms$. We discuss
below that evidence for a population of stars with rotation speeds
lagging the LSR by $\sim 100 \kms$ is also present in the metal-poor
star catalog of Beers et al (2000; hereafter B00), and in the
compilation of Gratton et al (2003, hereafter GCCLB).

In this {\it Letter} we reassess the status of Eggen's Arcturus group
by using modern distance, proper motions, radial velocities, and
metallicity determinations available in the literature. We argue that
the Arcturus group; the structure in the B00 and GCCLB compilations;
as well as the kinematically-odd population of stars discovered by
Gilmore, Wyse \& Norris (2002), are all part of the same, dynamically
coherent group of stars that might have originated in the disruption
of a fairly massive satellite early during the Galaxy's assembly.


\begin{figure*}[t]
\includegraphics[height=0.35\textheight,clip]{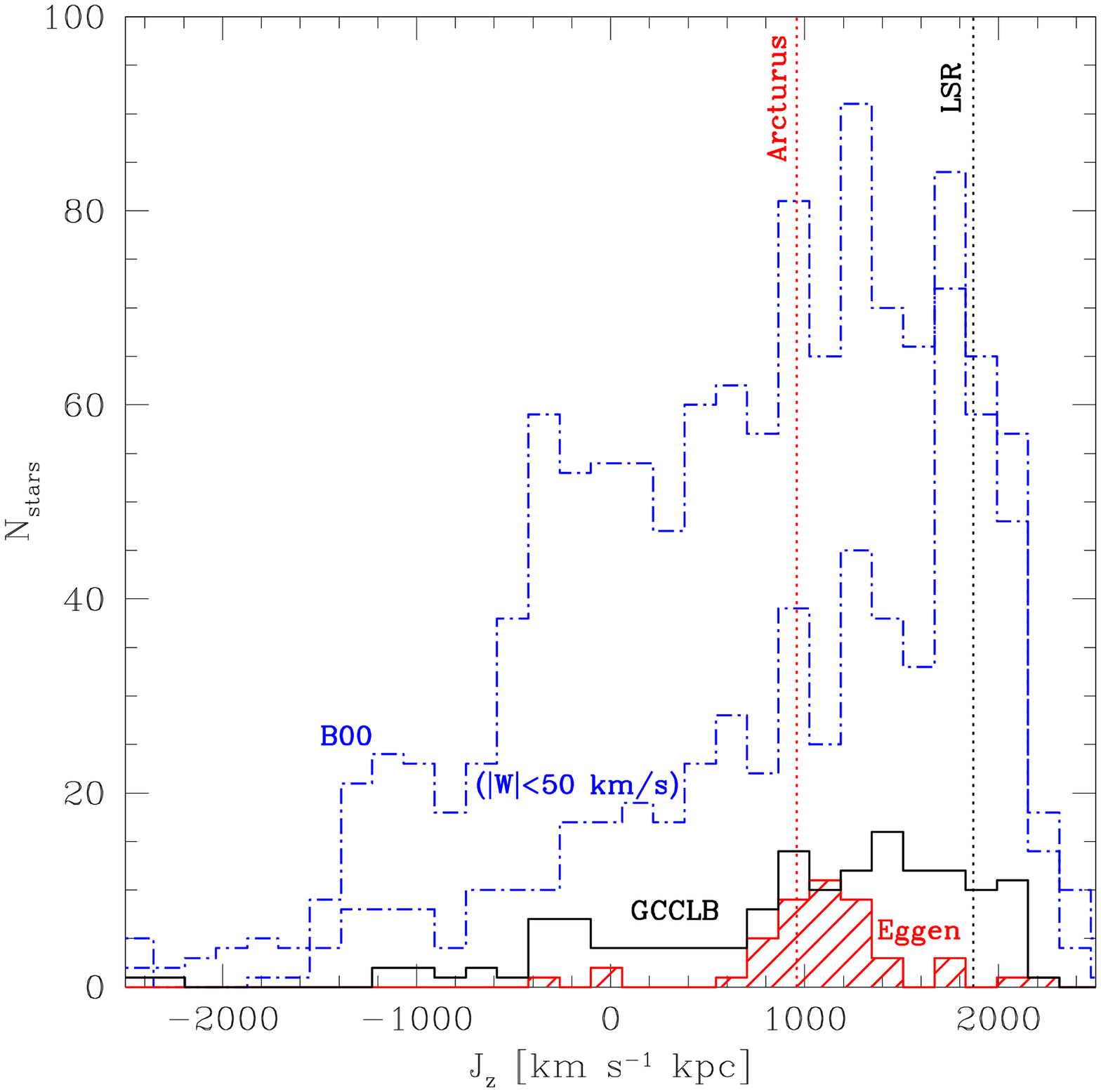}
\includegraphics[height=0.35\textheight,clip]{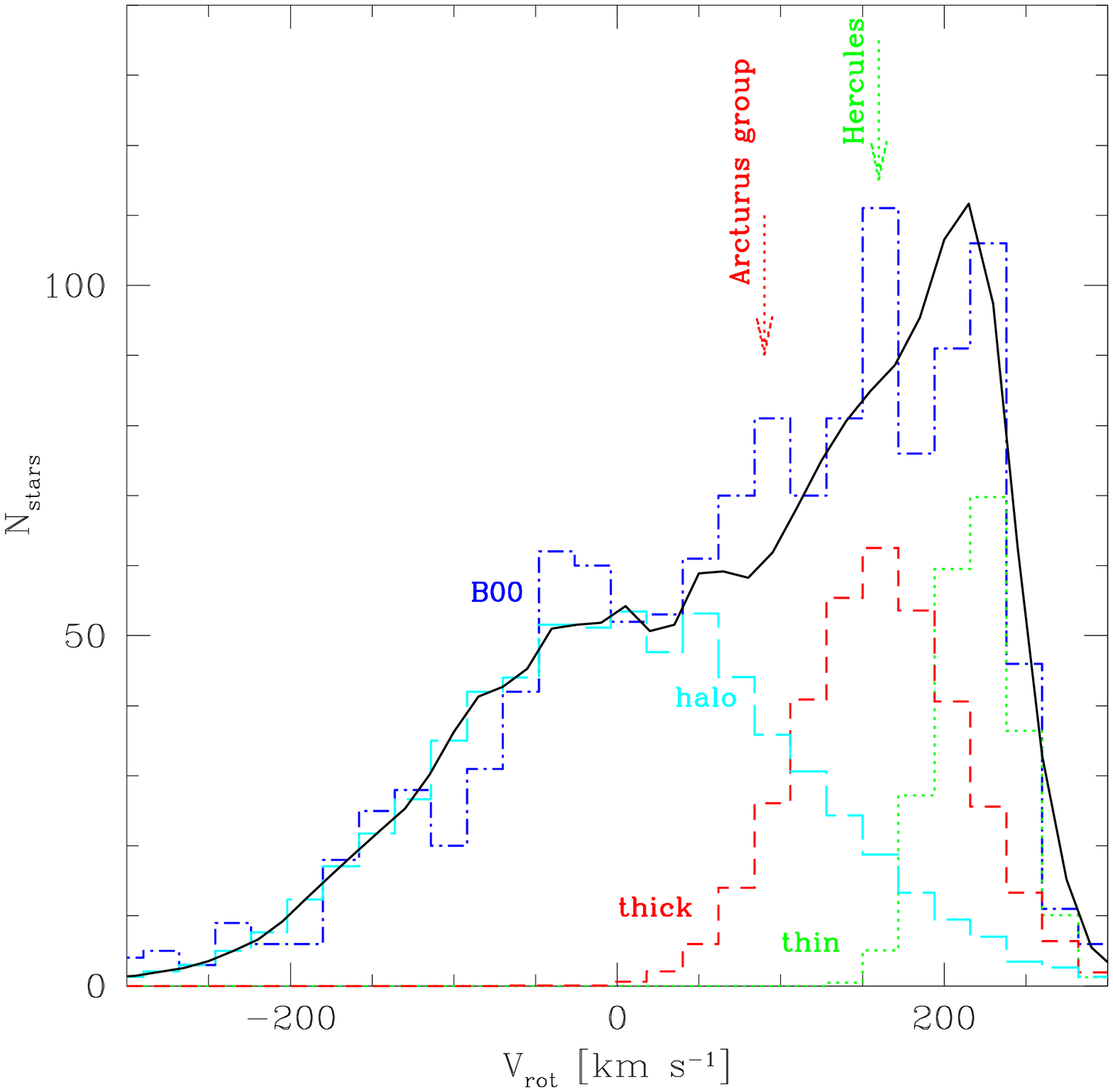}
\caption{Panel (a) shows the angular momentum distribution of stars in
the B00 catalog (top dot-dashed histogram); those in the Arcturus
group candidate list of Eggen (1996; 1998; shaded histogram at the
bottom); and stars in the compilation of Gratton et al. (2003; solid
histogram). The middle dot-dashed histogram correspond to all B00
stars on ``disk-like'' orbits, i.e., $|W|<50$ kms s$^{-1}$. There is
evidence in all these samples (which favor metal-poor stars) for an
``excess'' of stars with angular momentum coincident with the Arcturus
group at $J_z\sim 950$ $\kms$ kpc.  Panel (b) shows the rotational
velocity distribution of all stars in the B00 catalog, decomposed into
three canonical Galactic (Gaussian) components: halo, thick disk, and
thin disk. The solid curve denotes the sum of all three
components. Note the ``excess'' of stars lagging the LSR rotation by
$\sim 120 \kms$, coincident with the Arcturus group. See text for full
details.
\label{figs:jzhist}}
\end{figure*} 


\section{The Arcturus group}
\label{sec:arcts}

\subsection{Eggen's stars}
\label{ssec:eggens}

Over the years, Eggen compiled a list of stars kinematically similar
to Arcturus from catalogs of nearby stars with large proper motions
(Eggen 1971a; 1987; 1996; 1998). Distances to some of these stars,
however, were unavailable at the time, and Eggen would {\it assign}
them distances so that they had negligible velocity dispersion in $V$
($\sigma_V\sim 0.5$ $\kms$),
{\footnote{We use $U$, $V$, and $W$ to denote radial, tangential and vertical
velocities in Galactic coordinates, measured relative to the LSR. $U$ is
positive outwards, $V$ in the direction of the Sun's rotation, and $W$ towards
the North Galactic Pole. We assume that the Sun's ($U$,$V$,$W$) velocity is
($-9$,$12$,$7$) $\kms$.}}
as required by his interpretation. This somewhat arbitrary step has
made Eggen's interpretation controversial but, remarkably, his choice
led most candidates to lie along relatively narrow isochrones,
enabling Eggen to derive ages for these stars. He concluded that most
Arcturus group members are old ($\tau \gsim 10$ Gyrs) and metal poor
([Fe/H]$\sim-0.6$; Griffin \& Lynas-Gray, 1999).

The shaded histogram (labelled ``Eggen'') in Fig.~\ref{figs:jzhist}a shows the
distribution of the vertical component of the angular momentum, $J_z$, for stars
listed by Eggen (1996, 1998) as members of the Arcturus group. We have included
here only stars with available proper motions and Hipparcos parallaxes. This
effectively restricts Eggen's sample to $46$ stars within a sphere of radius
$\sim 300$ pc centered on the Sun. The dispersion in $J_z$ (corresponding to a
dispersion in $V$ of $\sigma_V\sim 50 \kms$) is much larger than demanded by
Eggen's interpretation of this association as a dissolving star cluster. Still,
many of the candidates are indeed on orbits that lag the Sun's rotation speed by
$\sim 100 \kms$.
Furthermore, many of these stars are on disk-like orbits; $\sim 70\%$
of them have vertical speeds that do not exceed $50 \kms$ and do not
climb farther than $\sim 1$ kpc out of the Galactic plane.  These
stars show no obvious net motion in the radial direction.

\subsection{Metal-deficient stars in the solar neighborhood}

B00 provide a catalog of metal-deficient stars compiled with the
specific aim of minimizing kinematic selection biases. This is the
largest sample of metal-poor ([Fe/H] $\le -0.6$) stars with available
abundances, distances, and radial velocities. Proper motions are also
available for many of these stars, making it a good sample to search
for substructure in the orbits of metal-deficient stars (see, e.g.,
Chiba \& Beers 2000; Brook et al. 2003).

The top dot-dashed histogram in Fig.~\ref{figs:jzhist}a shows the
angular momentum distribution for all stars in the B00 sample. The
full sample is dominated by the canonical ``thick disk'' component,
whose $J_z$ distribution peaks at $\sim 1300 \kms$ kpc, and lags the
LSR rotation speed by $50$--$60$ $\kms$.  The halo component
contributes a significant number of counter-rotating stars, and is
fairly prominent in this sample biased towards metal-poor stars.

The B00 distribution shows a number of prominent features that persist
even when the sample is restricted to stars with disk-like kinematics
($|W| \le 50 \kms$, see the lower dot-dashed histogram in
Fig.~\ref{figs:jzhist}a). We shall focus on two of these features,
also seen in the distribution of rotation-speeds shown in
Fig.~\ref{figs:jzhist}b ($V_{\rm rot}=J_z/R$, where $R$ is the polar radial
distance measured in the meridional plane).

The first feature, coincident with the peak velocity expected for the canonical
thick disk, likely signals the ``Hercules'' stream (Eggen 1971b). This stream
has $V \sim - 60 \kms$ and $U \sim 35$ $\kms$ (Skuljan et al. 1999), and is
thought to originate from dynamical resonances induced by the Galactic bar
(Dehnen 2000, Fux 2001). Intriguingly, the second feature, at $J_z\sim 950$
$\kms$ kpc, has an angular momentum similar to that of Arcturus. The same
feature is also seen in the GCCLB compilation of nearby metal-poor stars (see
solid histogram in Figure~\ref{figs:jzhist}a).

The statistical significance of this feature is difficult to assess,
since the sample is subject to a number of subtle and complex biases
and selection effects. We can nevertheless estimate it roughly by
comparing the actual number of stars with velocities similar to that
of Arcturus, with that expected for a random sample drawn from the
standard Galactic components: thin disk, thick disk and stellar
halo. This is shown in Fig.~\ref{figs:jzhist}b, where the $V_{\rm
rot}$ distribution has been decomposed into three Gaussians. From left
to right, the halo, thick disk, and thin disk components are assumed
to have (${\bar V}_{\rm rot}$, $\sigma$) equal to ($0$,$110$),
($160$,$50$) and ($220$,$25$) $\kms$, respectively.

The halo distribution is normalized to match the number of
counter-rotating stars, whereas the relative contributions of the
thick and thin disks are chosen to match the $V_{\rm rot}$
distribution of the remaining co-rotating stars. The fit obtained is
shown by the thick solid curve in Fig.~\ref{figs:jzhist}b. Most random
realizations are unable to account for the ``excess'' of stars
labelled ``Arcturus group''; only $16$ out of $1000$ realizations
give a number of stars comparable to what is observed.
Such odds seem compelling enough to warrant exploring further the provenance of
stars with such peculiar kinematics.

\subsection{The origin of the Arcturus group}
\label{ssec:extrag}

Could the Arcturus group be the result of perturbations to the orbits
of disk stars induced by the Galactic bar? As discussed above,
detailed modeling of such perturbations have linked stellar groupings
in phase space to orbital resonances in a rotating barred potential
(see, e.g., Dehnen 2000, Fux 2001). However, these perturbations
typically induce small velocity changes ($\sim 20$--$50$ $\kms$), and
therefore are unlikely to be responsible for the $110$--$130$ $\kms$
lag of stars in the Arcturus group. Moreover, such perturbations tend
to produce a net mean radial velocity with respect to the LSR, which
is not observed neither for the B00 candidates nor for stars in the
original Eggen's compilation (see Fig.~\ref{fig:vel}).

\subsubsection{The group as tidal debris}

A compelling alternative is that Arcturus is part of the tidal debris of a
disrupted satellite, akin to the interpretation of the SDSS ``ring'' proposed by
Helmi et al (2003). Intriguingly, the angular momentum of Arcturus is roughly
half of that of a circular orbit at the solar circle, reminiscent of the
dynamics of the SDSS ``ring''.
A further similarity with the outer Galactic ``ring'' is the lack of net radial
motion, which suggests that these structures have become recognizable as stars
of similar energy crowd at the apocenter of their orbits.  As discussed below,
an extragalactic origin for the Arcturus group would explain its sizeable
velocity dispersion, as well as its vertical extension and distinct metal
abundance.

We can test the plausibility of this scenario by following the disruption of a
small satellite with orbit similar to that of Arcturus in the potential of the
Galaxy. To this aim, we have integrated Arcturus' orbit $8$ Gyrs back in time
and use those orbital parameters to release a $10^8 \sm$ satellite (modelled
with $10^5$ particles after a King sphere with core and tidal radius $233$ pc
and $1.57$ kpc, respectively). The numerical simulations include the
self-gravity of the satellite using a quadrupole expansion of the internal
potential (Zaritsky \& White 1988), but are otherwise run in a fixed Galactic
potential, of the form given by Johnston et al. (1996). We shall focus on
particles (``stars'') located in a small volume around the Solar neighborhood, 5
to 8 Gyr after the satellite is disrupted.

\begin{figure}[t]
\includegraphics[height=0.3\textheight,clip]{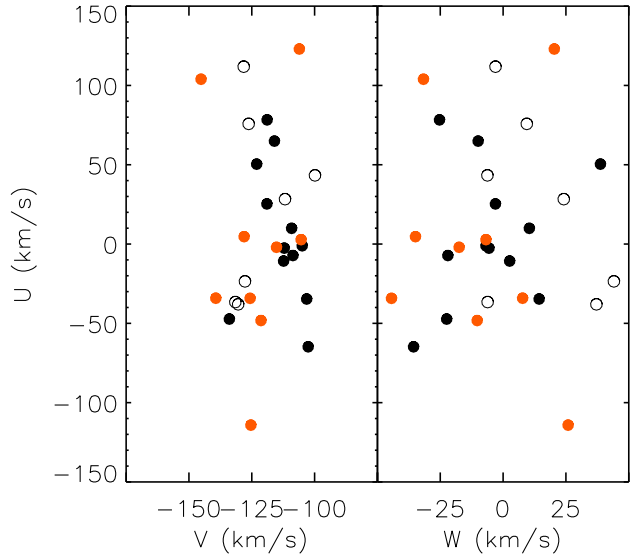}
\caption{Velocity components of candidate stars in the Arcturus group,
from Eggen's (black) and GCCLB (gray/red) compilations. Stars are selected as
candidates if $|V_{\rm rot} - V_{\rm Arct}| \le 20 \kms$ and $|W|
\le 50 \kms$. Open circles show the velocities of particles in the simulations located
in a small volume around the ``Sun'', 8 Gyr after infall. \label{fig:vel}}
\end{figure}

The open circles in Fig.~\ref{fig:vel} show the velocities of particles located
within 500 pc of the ``Sun'' (assumed to be at 8.5 kpc from the center of the
Galaxy, and on the Galactic plane), which should be compared to stream
candidates taken from the Eggen (solid black circles) and GCCLB (solid gray/red
circles) samples. The good agreement between the mean velocities of the
candidate stars and particles in the simulation indicate that the dynamics of
the stream is consistent with that expected for the debris of a shredded
satellite. Clues to the original mass of the satellite are preserved in the
velocity dispersion of the group, but will depend on the number of streams
expected to run through the solar neighbourhood at present. This is very hard to
ascertain given current uncertainties in the velocities.

\subsubsection{Chemical properties of the Arcturus group}

Supporting evidence for the ``shredded-satellite'' interpretation may be found
in the chemical properties of stars in the Arcturus group. The Beers et al
(2000) catalog lists [Fe/H] estimates for all stars, and further examination
shows that stars with angular momentum consistent with the Arcturus group span a
wide range of metallicities, roughly $-2.5<$[Fe/H]$<-0.5$.  Does this broad
metallicity distribution argue against Eggen's suggestion that these stars are
part of a disrupting cluster born out of the same molecular cloud ? It is indeed
clear that most of the B00 stars in the appropriate $J_z$ range are metal-poor,
but we would expect to find a significant contamination from halo and metal-poor
thick disk stars in this range of $J_z$ (c.f. Fig~\ref{figs:jzhist}b).

On the other hand, the tightness of the [$\alpha$/Fe] relation for the group
candidates in Fig.~\ref{figs:afeh}, compared with the non-group candidates, does
suggest a common brief star formation history for the stars in this range of
$J_z$. This is best shown using the GCCLB dataset, which includes abundance
estimates for elements other than Fe. This dataset may thus be used to search
for potential features in the chemical properties of group candidate stars. In
Figure~\ref{figs:afeh}, open symbols are used for all stars in the GCCLB sample,
whilst filled circles are used for Arcturus group candidates; i.e., stars with
$J_z$ in the range $(700,1100)$ $\kms$ kpc, and $|W|<50$ $\kms$. Candidate stars
of the group define a tight sequence in the [$\alpha$/Fe]--[Fe/H] plane, much
tighter than expected for stars of similar [Fe/H] selected at random from the
full GCCLB sample.

Interestingly, the group candidates appear to follow, with minimal scatter, a
well defined enrichment pattern roughly consistent with simple self-enrichment
models, where the overabundance in [$\alpha$/Fe] (relative to solar) suggests a
gas consumption timescale short compared with that of supernova type Ia
ignition. The curve in Fig.~\ref{figs:afeh} shows the predictions of Matteucci
\& Francois (1989) for a one-zone model{\footnote{The curve corresponds to
[Si/Fe] vs [Fe/H] in their model 1. Results are similar for other $\alpha$
elements.}} that self-enriches to [Fe/H]$=-1$ in about $1$ Gyr, before the
Fe-rich yield of SNIa start to play a role, leading to a sudden change in the
slope of the curve. The good agreement between the model prediction and the
candidate stars should be taken as a demonstration of the plausibility of our
interpretation rather than as an endorsement of this particular model; other
choices of model parameters may lead to similarly good fits, although all are
likely to require a short timescale for enrichment.

We conclude that the coherence in the dynamical and chemical
properties of stars in the group suggest that they originate from a
single satellite system that was dragged and shredded into the disk of
the Galaxy early during its assembly.

\section{Summary and Discussion}
\label{sec:disc}

The peculiar angular momentum of the Arcturus group, together with its
moderate velocity dispersion and singular metal abundance, all point
to a possible extragalactic origin for the group.  Age estimates are
available for some stars in the group (Fuhrmann 2000, 2003), and
suggest that it is perhaps as old as $10$-$12$ Gyr. This would imply
that the group is the debris from an ancient accretion event.

This accretion event has probably contributed a significant number of
metal-deficient stars to the solar neighborhood, as shown by the
prominence of the group in samples that favor metal-deficient
stars. The presence of this dynamically-coherent group may explain
some of the variance in the results of dynamical studies of the thick
disk. Indeed, some of these studies have concluded that the bulk of
the thick disk lags the LSR by only $\sim 30$--$50 \kms$ (Beers \&
Sommer-Larsen 1995, Ojha et al 1996, Norris 1999), whereas others have
favored a lag of $\sim 100 \kms$ instead (Wyse \& Gilmore 1986, Chen
et al 2000, Fuhrmann 2000, 2003), possibly reflecting the different
contribution of stars in the group to each sample.

\begin{figure}[t]
\begin{center}
\includegraphics[height=0.35\textheight,clip]{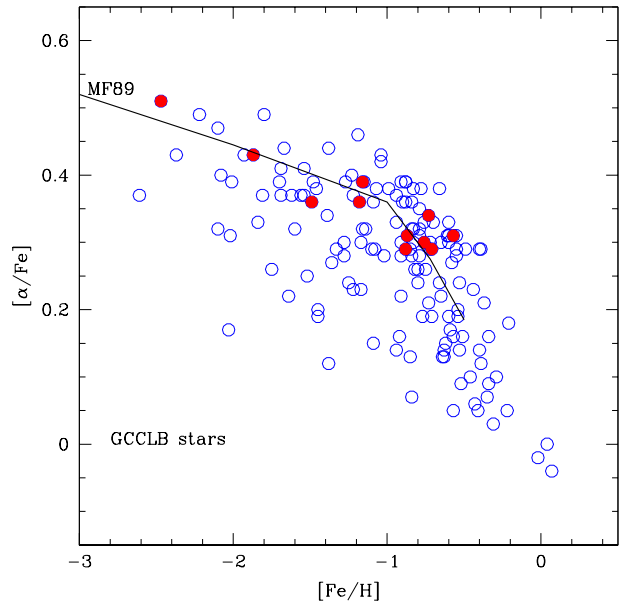}
\end{center}
\caption{Abundance ratios for stars in the GCCLB sample. Open circles
correspond to all stars in each sample, whereas filled circles indicate
candidate stars of the Arcturus group. Note that these stars delineate a well
define sequence in the abundance ratios that may be interpreted as resulting
from a short episode of star formation where a system self-enriched to a
metallicity of order one-third solar. The curve correspond to model 1 of
Matteucci \& Francois (1989). See text for further details.
\label{figs:afeh}}
\end{figure}

There are a number of inquiries that may further validate (or refute)
our interpretation. For example, confirming the link between the
Arcturus group and the debris identified by Gilmore, Wyse \& Norris
outside the plane would help to uphold the group's extragalactic
origin. Most of the stars in the Arcturus group are located close to
their orbital apocenter, where density enhancements are expected: we
may have detected a ``shell'' similar to those observed around
elliptical galaxies (Helmi et al 2003). Other distinct features of
this shell interpretation would include a rapid decline in the
importance of the group at increasingly large Galactocentric distance
(shells are ``sharp'') and, of course, the presence of other shells at
different Galactocentric radii.  It may also be possible to find other
members of the progenitor of the Arcturus group by searching for stars
on disk-like eccentric orbits with metal abundances consistent with
the ``sequence'' outlined by the group in Figure~\ref{figs:afeh}.

Finally, it is oddly gratifying to think of stars visible to the naked
eye, such as Arcturus, as silent night-sky witnesses of the merging
history of the Milky Way.

\acknowledgements
We thank Brad Gibson and the hospitality of Swinburne University of
Technology, where this work was conceived. JFN is supported by several
grants from Canada's NSERC and from the Canadian Foundation for
Innovation.

\end{document}